**Large-scale quantum-emitter arrays in atomically thin semiconductors**

Carmen Palacios-Berraquero[†1], Dhiren M. Kara[†1], Alejandro R.-P. Montblanch[1], Matteo Barbone[1,2], Pawel Latawiec[3], Duhee Yoon[2], Anna K. Ott[2], Marko Loncar[3], Andrea C. Ferrari[2*] and Mete Atatüre[1*]

[1]*Cavendish Laboratory, University of Cambridge, JJ Thomson Ave., Cambridge CB3 0HE, UK*

[2]*Cambridge Graphene Centre, University of Cambridge, Cambridge CB3 0FA, UK*

[3]*John A. Paulson School of Engineering and Applied Science, Harvard University, 29 Oxford Street, Cambridge MA 02138, USA*

[†]These authors contributed equally to this work.

**The flourishing field of two-dimensional (2D) nanophotonics[1–4] has generated much excitement in the quantum technologies community after the identification of quantum emitters (QEs) in layered materials (LMs)[5–10]. LMs offer many advantages as platforms for quantum circuits[4,11], such as integration within hybrid technologies[11], valley degree of freedom[12–14] and strong spin-orbit coupling[15]. QEs in LMs, however, suffer from uncontrolled occurrences, added to the uncertainty over their origin, which has been linked to defects[5–10] and strain gradients[16–18]. Here, we report a scalable method to create arrays of single-photon emitting QEs in tungsten diselenide (WSe$_2$) and tungsten disulphide (WS$_2$) using a nanopatterned silica substrate. We obtain devices with QE numbers in the range of hundreds, limited only by the flake size, and a QE yield approaching unity. The overall quality of these deterministic QEs surpasses that of their randomly appearing counterparts, with spectral wanderings of ~0.1 meV – an order of magnitude lower than previous reports[5–9]. Our technique solves the scalability challenge for LM-based quantum photonic devices.**

In order to create large-scale QE arrays in LMs, we subject the active material to patterned structures fabricated on the substrate in order to create spatially localised physical disturbances to the otherwise flat LM flakes. To this end, we first pattern arrays of nanopillars of different heights, ranging from 60 to 190 nm, on silica substrates using electron beam lithography. Figure 1a shows a scanning electron microscope (SEM) image of one such substrate of 130-nm nanopillar height. We place layers of $WSe_2$ and $WS_2$ on the nanopillars as follows. Bulk $WSe_2$ and $WS_2$ crystals are characterized prior to exfoliation as described in our previous work[19]. These are then exfoliated on a polydimethylsiloxane (PDMS) layer by micromechanical cleavage[20,21]. Single layer (1L) samples are identified first by optical contrast[22], and the selected 1L-$WSe_2$ and 1L-$WS_2$ flakes are then placed onto the patterned nanopillar substrate via an all-dry viscoelastic transfer technique due to their higher adhesion to $SiO_2$[23], as schematically shown in Fig. 1b. After exfoliation and transfer, the 1L-$WSe_2$ and 1L-$WS_2$ flakes are characterized by Raman spectroscopy[24,25], photoluminescence (PL)[26] and atomic force microscopy (AFM), confirming the transfer and that the process does not damage the samples (see Supplementary Section 1 for the corresponding spectra and discussion). Figure 1c is an AFM scan of a 1L-$WSe_2$ flake over a single nanopillar. The bottom panel of Fig. 1c plots the height profile of the 1L-$WSe_2$ flake taken along the dashed pink line. This reveals how the flake (solid pink line) tents over the nanopillar. The blue-shaded area corresponds to the measured profile of a bare nanopillar. Figure 1d is a dark field optical microscopy (DFM) image of part of a 43,000 $\mu m^2$ 1L-$WSe_2$ flake on a substrate patterned with a 4-$\mu m$-spaced nanopillar array with nominal height of 130 nm. The regularly spaced bright spots correspond to nanopillar sites. We see locations providing brighter scattering (two examples are encircled in pink) and others showing fainter intensity (two examples are encircled in blue). By correlating with AFM measurements we find that the former correspond to locations where the 1L-$WSe_2$ tents over the nanopillars and the latter correspond to locations where the flake is pierced by the nanopillars (see Supplementary Section 2). On average, we find that 2/3 of the sites are not pierced during the deposition step.

Figure 2a is an integrated raster scan map of PL emission at ~10 K of six adjacent non-pierced nanopillar sites in the region enclosed by the green dashed line in Fig. 1d. The most prominent feature is the ~×10 increase in intensity at the location of every nanopillar. Figure 2b reveals the source of this emission intensity enhancement: spectra taken at each nanopillar location display bright sub-nanometre linewidth emission peaks. Figure 2c demonstrates the single-photon nature of this emission via photon-correlation measurements taken (from left to right) at the first, third and fourth nanopillar locations. 10-nm bandpass filters, indicated by the pink, green and blue highlighted areas in the panels of Fig. 2b, select the spectral windows for the photon-correlation measurements. We obtain $g^{(2)}(0)$ values of

0.0868 ± 0.0645, 0.170 ± 0.021 and 0.182 ± 0.028, respectively, uncorrected for background emission or detector response. While these values already surpass those in previous reports[5–9], we expect the quality of the single-photon emission from the QEs to improve under resonant excitation[27]. Of the 53 unpierced nanopillar sites in this substrate we found sub-nm emission peaks in 51 of them giving ~96% yield in QE generation. Their emission wavelength ranges between 730 nm and 820 nm (see Supplementary Section 3 for statistics), equivalent to a redshift distributed between 50 and 280 meV from the unbound exciton emission energy at ~1.755 eV[28], as observed for the naturally occurring QEs in $WSe_2$[5–9]. The fine-structure splitting (200-730 µeV) and the emission linewidths as narrow as ~180 µeV (~0.08 nm) are also consistent with previous reports[5–9] (see Supplementary Section 3) advocating that these deterministically created QEs are of the same nature as the randomly appearing ones.

To study the effect of nanopillar height, we carry out similar optical measurements of 1L-$WSe_2$ flakes deposited on nanopillars of height ~60 nm and ~190 nm. The spectra taken at the 60-nm nanopillars have multiple peaks of ~1-nm linewidth on average (see Supplementary Section 3 for example spectra). In contrast, Fig. 2d is a representative spectrum taken from the 190-nm nanopillars, displaying a better isolated, single sub-nm emission peak. The inset reveals a 722-µeV fine-structure splitting for this QE. We do not see clear nanopillar height dependence in the emission wavelength and fine-structure splitting (see Supplementary Section 3 for statistics). However, increasing the nanopillar height does reduce the spread in the number of peaks arising at each location. We verify this trend in Fig. 2e, a histogram of the probability that a given number of sub-nm emission peaks appear per nanopillar, for the different nanopillar heights (60, 130 and 190 nm in white, blue and purple, respectively). The likelihood of creating a single QE grows as nanopillar height is increased. For the 190-nm nanopillars, 50% of all nanopillar sites host a single QE with one emission peak, as indicated by the purple bars. Spectral wandering of the peaks as a function of time also displays a strong dependence on the nanopillar height. To quantify this dependence, we record the maximum range of emission wavelength wandering per QE over tens of seconds. The solid black circles in Fig. 2f correspond to the mean of these values for each group of QEs pertaining to each nanopillar height, for 17 different QEs in total, with the error bars displaying the variance of these distributions. We observe a reduction from a few meV for 60-nm height nanopillars to below 0.25 meV (average) for the tallest 190-nm nanopillars (see Supplementary Section 4), reaching as low as 0.1 meV. To the best of our knowledge, this is the lowest spectral wandering seen in LM QEs[5–9]. Hence, these deterministic QEs are comparable, and even superior, in spectral stability to their randomly appearing counterparts. The dependence of certain QE characteristics on nanopillar height, along with shifts in the delocalised neutral exciton peak ($X^0$) at room temperature[26] at the nanopillar locations (see Supplementary Section 5), suggest that a localised strain gradient induced by the nanopillars may be playing an active role in producing QEs, as well as determining their specific optical properties[16–18].

The method we present for QE creation is not restricted to a specific LM. We predict a similar effect on different LMs and test this by using 1L-$WS_2$. Figure 3a shows an integrated PL intensity raster scan map taken at ~10 K of a 1L-$WS_2$ on a substrate with 170-nm-high nanopillars square array spaced by 3 μm. The inset shows a true-colour DFM image of the same flake, where the red areas (due to fluorescence) are 1L-$WS_2$. Once again, the brighter spots correspond to the unpierced nanopillar locations, as verified by AFM measurements, and show clear overlap with the bright fluorescence spots in the PL intensity image where, similar to $WSe_2$, intensity is increased (here by a factor ~4) at every one of the 22 non-pierced nanopillar sites in the flake. Panel 1 of Fig. 3b shows the typical 1L-$WS_2$ emission spectrum at ~10 K[19], measured from a flat region of the same flake away from the nanopillars. The $X^0$ and $X^-$ unbound excitons are labelled in the figure, while the broad red-shifted emission band arises from weakly localised or defect-related excitons in the monolayer at low temperatures[19], and is present in this material regardless of location. Panels 2 and 3 of Fig. 3b show representative PL emission spectra taken at nanopillars of heights ~170 and ~190 nm, respectively, where once again sub-nm spectral features arise. We also note that we observe fine-structure splitting for $WS_2$ in these QEs, which range from 300 to 810 μeV (see Supplementary Section 6), as represented in the panel insets corresponding to the spectral regions highlighted in red. We also measure the spectrum of several $WS_2$ QEs as a function of time (see Supplementary Section 6) and find all spectral wandering values below 0.5 meV over 1-2 minutes. Figure 3c shows statistics on QE emission wavelength collected for over ~80 QEs for 1L-$WS_2$ on 170-nm (white bars) and 190-nm (red bars) nanopillars. The wavelength distribution of the sub-nm emission lines, typically in the 610-680 nm region (53-300 meV redshift from $X^0$)[19], is as narrow as ~20 nm for the 190-nm nanopillars. Most nanopillar sites on $WS_2$ show multiple sub-nm lines, suggesting the creation of several QEs at each site for these nanopillar heights. Figure 3d plots a histogram of the number of sub-nm peaks appearing at each nanopillar for both nanopillar heights. The trend is similar to that seen to WSe$_2$, where higher nanopillars lead to a narrower spread in the number of peaks towards a higher likelihood of creating a single QE at each nanopillar site. We note that we obtain a 95% yield of QE creation in 1L-WS$_2$ on non-pierced nanopillars. Further, ~75% of these display two or less sub-nm emission peaks. In contrast, the 60-nm and the 130-nm-high nanopillars do not result in any QE occurrence (see Supplementary Section 7 for examples of these measurements). This strong dependence of QE creation on nanopillar height further points towards a potentially critical role played by local strain. Despite previous efforts to measure QEs in 1L-WS$_2$, there has only been one previous report of single-photon emission in this material[19]. These results suggest that the rarity of QEs in exfoliated WS$_2$ flakes on flat substrates might indeed be due to the lack of sufficient deformation, provided here by tall nanopillars.

We presented a simple method for the deterministic creation of scalable arrays of quantum-light emitters embedded in LMs emitting at different regions of the optical spectrum. The reliability of the technique will accelerate experimental studies of QEs in TMDs, which at present rely on their rather rare and random occurrence[5–9]. In the immediate future, a detailed study is necessary in order to achieve a better understanding of the specific role of nanopillar height and geometry in defining the characteristics of the quantum emission. We expect tunability of the optical emission by varying the shapes of the underlying nanostructures. In this respect, interesting possibilities to realise dynamical circuits using micro-electro-mechanical systems and piezoelectric tuning exist. Heterostructures may enable new routes towards tunnel-coupled quantum devices and the formation of QE molecules. Several approaches are being investigated for the production of wafer-scale samples[29,30], which could lead to rapid optimisation. While our approach is already compatible with standard silicon processing techniques, it is nevertheless not restricted to the specific properties of the substrate. In fact, even nanodiamonds of the appropriate dimensions, drop-casted onto silica substrates, are able to create QEs in 1L-WSe$_2$ (see Supplementary Section 8). The flexibility in the choice of substrate, in turn, provides an opportunity to create hybrid quantum devices where LM QEs can be coupled to quantum systems in other materials such as spins in diamond and silicon carbide.

## Methods

**Substrate Preparation:** The silica nanopillar substrate is fabricated with a high-resolution direct-write lithographic process via spin-on-glass polymer hydrogen silsesquioxane (HSQ)[31]. First, a wafer with 2 μm thermal oxide is cleaved and then cleaned. HSQ resist (FOx-16, Dow-Corning) is diluted with methyl isobutyl ketone (MIBK) in different ratios and spun onto the substrate, giving variable thickness depending on the dilution. After baking at 90 °C for 5 minutes, the substrate is exposed in an electron beam lithography tool (Elionix F-125) and then developed in a 25% solution of tetramethyl ammonium hydroxide (TMAH) developer and rinsed in methanol. To convert the defined structures into pure $SiO_2$, we apply rapid thermal annealing at 1000 °C in an oxygen atmosphere [32], resulting in arrays of sharply-defined sub-100 nm silica nanopillars.

**Optical Measurements:** Room temperature Raman and PL measurements are carried out using a Horiba LabRam HR Evolution microspectrometre equipped with a ×100 objective (numerical aperture 0.9) and a spot size <1μm. The pixel-to-pixel spectral resolution for the Raman measurements is ~0.5 $cm^{-1}$. Bragg gratings (BraggGrate) are used to detect the ultralow frequency Raman peaks. The power is kept below 50 μW to prevent heating effects. The excitation wavelength used is 514.5 nm for $WSe_2$ and 457 nm for $WS_2$.

A variable-temperature helium flow cryostat (Oxford Instruments Microstat HiRes2) is used to perform low-temperature PL measurements. These PL measurements are performed using a home-built confocal microscope mounted on a three-axis stage (Physik Instrumente M-405DG) with a 5-cm travel range, 200-nm resolution for coarse alignment and a piezo scanning mirror (Physik Instrumente S-334) for high-resolution raster scans. PL is collected using a 1.7-mm working distance objective with a numerical aperture of 0.7 (Nikon S Plan Fluor ×60) and detected on a fibre-coupled single-photon-counting module (PerkinElmer: SPCM-AQRH). Photon correlations from a Hanbury Brown and Twiss interferometer are recorded with a time-to-digital converter (quTAU). A double grating spectrometer (Princeton Instruments) is used for acquiring spectra. For PL measurements, the excitation laser (532 nm, Thorlabs MCLS1) is suppressed with a long pass filter (550 nm Thorlabs FEL0550).

**Acknowledgements**

We acknowledge financial support from the Marie Skłodowska-Curie Actions Spin-NANO, Grant No. 676108, EU Graphene Flagship, ERC Grants Hetero2D and PHOENICS, EPSRC Grants EP/K01711X/1, EP/K017144/1, EP/N010345/1, EP/M507799/1, EP/L016087/1, Quantum Technology Hub NQIT EP/M013243/1, the EPSRC Cambridge NanoDTC, Graphene Technology CDT, EP/G037221/1, and the STC Center for Integrated Quantum Materials (NSF Grant No. DMR-1231319). We would like to thank H. S. Knowles and P. Borisova for technical assistance.

**References**

1. Bonaccorso, F., Sun, Z., Hasan, T. & Ferrari, A. C. Graphene photonics and optoelectronics. *Nat. Photonics* **4,** 611–622 (2010).
2. Koppens, F. H. L. *et al.* Photodetectors based on graphene, other two-dimensional materials and hybrid systems. *Nat. Nanotechnol.* **9,** 780–793 (2014).
3. Ferrari, A. C. Science and technology roadmap for graphene, related two-dimensional crystals, and hybrid systems. *Nanoscale* **7,** 4598–4810 (2014).
4. Xia, F., Wang, H., Xiao, D., Dubey, M. & Ramasubramaniam, A. Two-dimensional material nanophotonics. *Nat. Photonics* **8,** 899–907 (2014).
5. Tonndorf, P. *et al.* Single-photon emission from localized excitons in an atomically thin semiconductor. *Optica* **2,** 347 (2015).
6. Srivastava, A. *et al.* Optically active quantum dots in monolayer $WSe_2$. *Nat. Nanotechnol.* **10,** 491–496 (2015).
7. He, Y.-M. *et al.* Single quantum emitters in monolayer semiconductors. *Nat. Nanotechnol.* **10,** 497–502 (2015).
8. Koperski, M. *et al.* Single photon emitters in exfoliated $WSe_2$ structures. *Nat. Nanotechnol.* **10,** 503–506 (2015).
9. Chakraborty, C., Kinnischtzke, L., Goodfellow, K. M., Beams, R. & Vamivakas, A. N. Voltage-controlled quantum light from an atomically thin semiconductor. *Nat. Nanotechnol.* **10,** 507–511 (2015).
10. Tran, T. T., Bray, K., Ford, M. J., Toth, M. & Aharonovich, I. Quantum emission from hexagonal boron nitride monolayers. *Nat. Nanotechnol.* **11,** 37–41 (2015).
11. Majumdar, A. *et al.* Hybrid 2D Material Nanophotonics: A Scalable Platform for Low-

Power Nonlinear and Quantum Optics. *ACS Photonics* **2,** 1160–1166 (2015).

12. Cao, T. *et al.* Valley-selective circular dichroism of monolayer molybdenum disulphide. *Nat. Commun.* **3,** 887 (2012).
13. Zeng, H., Dai, J., Yao, W., Xiao, D. & Cui, X. Valley polarization in $MoS_2$ monolayers by optical pumping. *Nat. Nanotechnol.* **7,** 490–493 (2012).
14. Mak, K. F., He, K., Shan, J. & Heinz, T. F. Control of valley polarization in monolayer $MoS_2$ by optical helicity. *Nat. Nanotechnol.* **7,** 494–498 (2012).
15. Kormányos, A., Zólyomi, V., Drummond, N. D. & Burkard, G. Spin-orbit coupling, quantum dots, and qubits in monolayer transition metal dichalcogenides. *Phys. Rev. X* **4,** 1–16 (2014).
16. Kumar, S., Kaczmarczyk, A. & Gerardot, B. D. Strain-Induced Spatial and Spectral Isolation of Quantum Emitters in Mono- and Bilayer $WSe_2$. *Nano Lett.* **15,** 7567–7573 (2015).
17. Branny, A. *et al.* Discrete quantum dot like emitters in monolayer $MoSe_2$: Spatial mapping, magneto-optics, and charge tuning. *Appl. Phys. Lett.* **108,** 142101 (2016).
18. Kern, J. *et al.* Nanoscale Positioning of Single-Photon Emitters in Atomically Thin $WSe_2$. *Adv. Mater.* (2016). doi:10.1002/adma.201600560
19. Palacios-Berraquero, C. et al. Atomically thin quantum light-emitting diodes. *Nat. Commun.* 7, 12978 doi: 10.1038/ncomms12978 (2016).
20. Novoselov, K. S. *et al.* Two-dimensional atomic crystals. *Proc. Natl. Acad. Sci. U. S. A.* **102,** 10451–3 (2005).
21. Bonaccorso, F. *et al.* Production and processing of graphene and 2d crystals. *Mater. Today* **15,** 564–589 (2012).
22. Casiraghi, C. *et al.* Rayleigh imaging of graphene and graphene layers. *Nano Lett.* **7,** 2711–2717 (2007).
23. Castellanos-Gomez, A. *et al.* Deterministic transfer of two-dimensional materials by all-dry viscoelastic stamping. *2D Mater.* **1,** 011002 (2014).
24. Terrones, H. *et al.* New first order Raman-active modes in few layered transition metal dichalcogenides. *Sci. Rep.* **4,** 4215 (2014).
25. Zhao, W. *et al.* Lattice dynamics in mono- and few-layer sheets of $WS_2$ and $WSe_2$. *Nanoscale* **5,** 9677–83 (2013).
26. Zhou, B. *et al.* Evolution of electronic structure in Atomically Thin Sheets of $WS_2$ and $WSe_2$. *ACS Nano* **7,** 791–797 (2013).
27. Kumar, S. *et al.* Resonant laser spectroscopy of localized excitons in monolayer $WSe_2$.

*Optica* **3,** 882 (2016).

28. Wang, G. *et al.* Valley dynamics probed through charged and neutral exciton emission in monolayer WSe$_2$. *Phys. Rev. B* **90,** (2014).
29. Kang, K. *et al.* High-mobility three-atom-thick semiconducting films with wafer-scale homogeneity. *Nature* **520,** 656–660 (2015).
30. Ling, X. *et al.* Parallel Stitching of 2D Materials. *Adv. Mater.* **28,** 2322–2329 (2016).
31. Grigorescu, A. E. *et al.* Resists for sub-20-nm electron beam lithography with a focus on HSQ: state of the art. *Nanotechnology* **20,** 292001 (2009).
32. Holzwarth, C. W., Barwicz, T. & Smith, H. I. Optimization of hydrogen silsesquioxane for photonic applications. *J. Vac. Sci. Technol. B Microelectron. Nanom. Struct.* **25,** 2658 (2007).

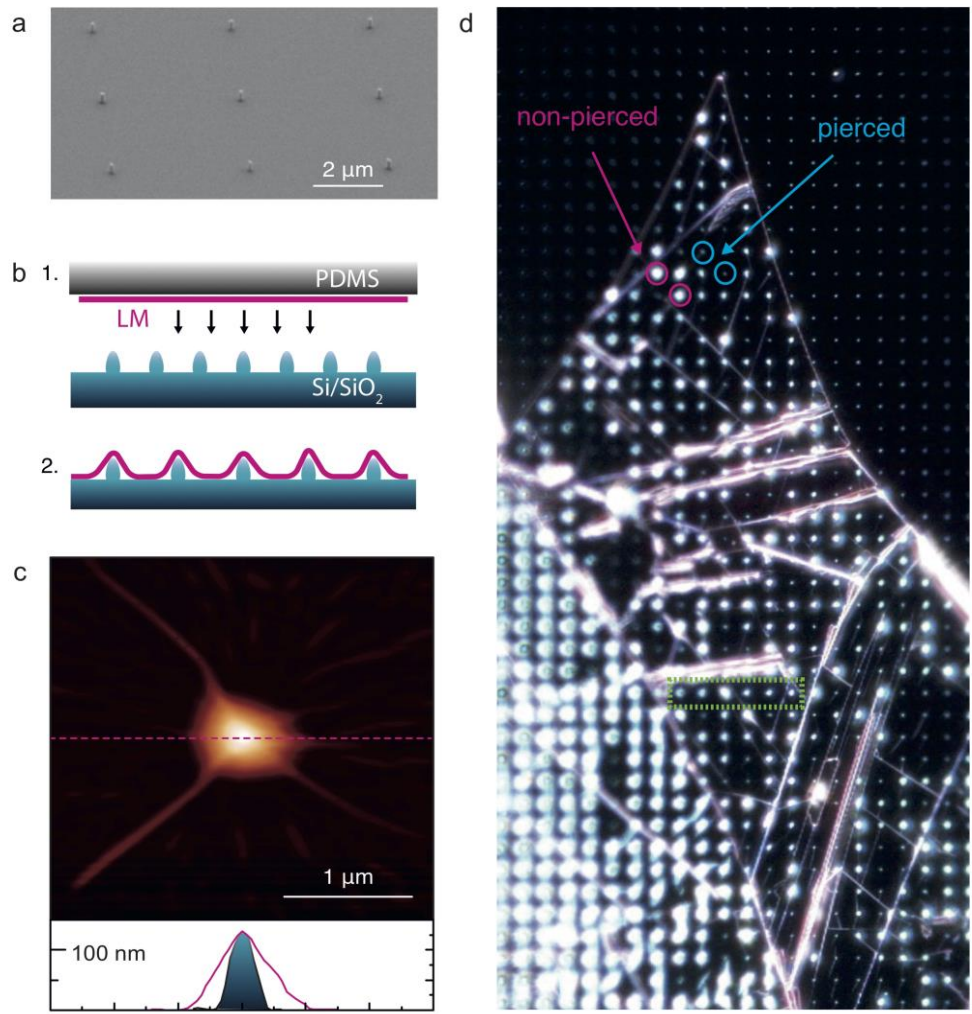

**Figure 1 | Scalable quantum confinement arrays: fabrication and characterisation. a,** SEM image of nanopillar substrate, fabricated by electron beam lithography. The black scale bar is 2 μm. **b,** Illustration of the fabrication method: 1. Mechanical exfoliation of LM on PDMS and all-dry viscoelastic deposition on patterned substrate. 2. Deposited LM on patterned substrate. **c,** Top panel shows an AFM scan of 1L-WSe$_2$ on a nanopillar. Bottom panel shows the AFM height profile of a bare nanopillar (blue-shaded region) and of the flake deposited over it (pink line), measured along the dashed pink line cut in the top panel. **d,** Dark field optical microscopy image (real colour) of 1L-WSe$_2$ on nanopillar substrate (130 nm high, 4-μm separation). The full image corresponds to a 170-μm by 210-μm area. The green box highlights six adjacent nanopillars within the 1L-WSe$_2$ region, measured in Fig. 2. The blue circles indicate two pierced nanopillars, and the pink circles indicate two non-pierced nanopillars.

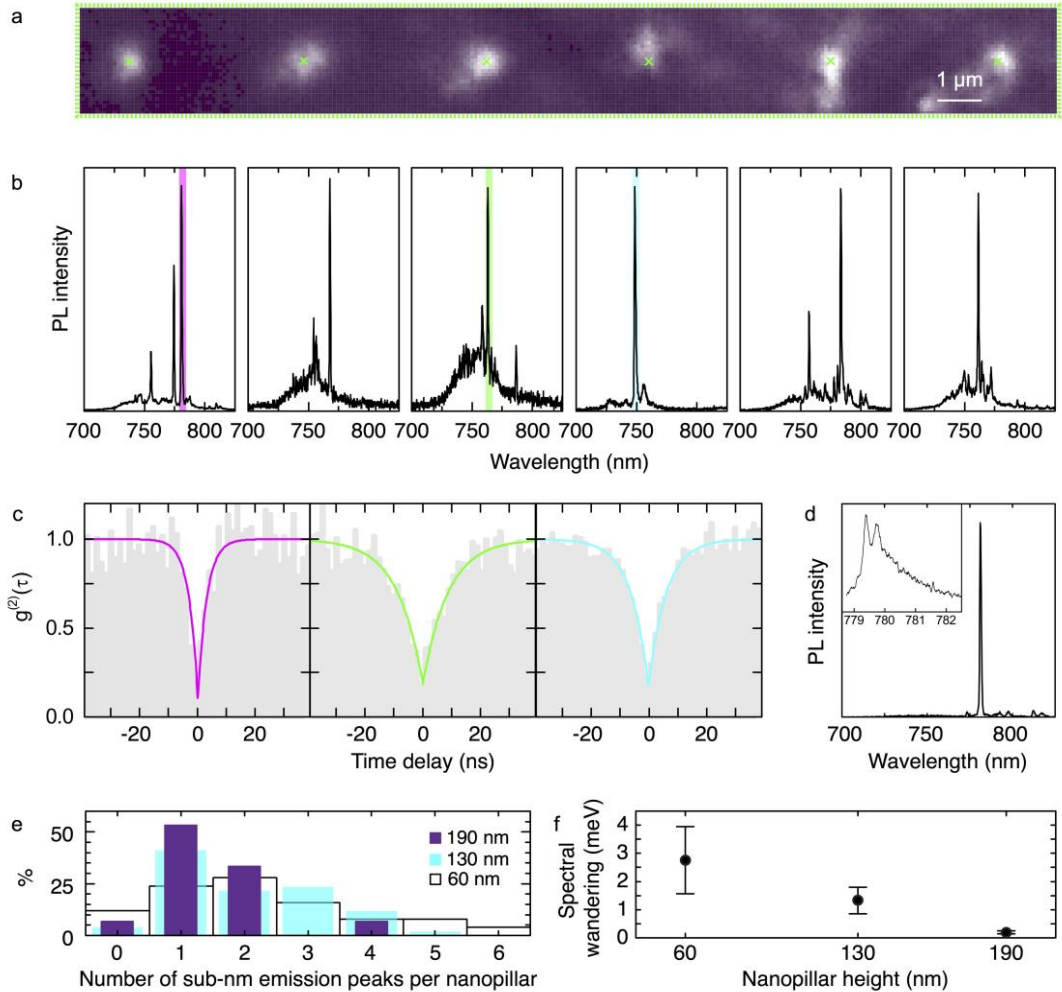

**Figure 2 | Creation of quantum emitter arrays in 1L-WSe$_2$. a,** Integrated PL intensity raster scan of the region enclosed by the green rectangle in Fig. 1d, taken under 200-nW/μm$^2$, 532-nm laser excitation at 10 K. Green crosses mark the position of the six nanopillars beneath the 1L-WSe$_2$. Colour-scale bar maximum is 160 kcounts/s. **b,** PL spectra taken at each of the corresponding green crosses in a, from left to right respectively, showing the presence of narrow lines at each nanopillar location. **c,** Photon correlation measurements corresponding to the filtered spectral regions (10 nm wide) enclosed by the blue, green and pink rectangles, in panel b, with $g^{(2)}(0)$=0.087 ± 0.065, 0.17 ± 0.02 and 0.18 ± 0.03, and rise times of 8.81 ± 0.80 ns, 6.15 ± 0.36 ns and 3.08 ± 0.41 ns, respectively. **d,** Spectrum taken from a 1L-WSe$_2$ on a 190-nm nanopillar, showing lower background and a single sub-nm emission peak. Higher resolution spectrum in the inset reveals the fine-structure splitting of this QE peak. **e,** Probability distribution (in %) of the number of emission lines per nanopillar

for samples using different nanopillar heights (60, 130 and 190 nm in white, blue and purple, respectively). A trend of higher probability of single QE emission peaks per nanopillar location with increasing height is evident, reaching 50% for 190-nm nanopillars. **f,** Increasing nanopillar height also leads to a reduction of spectral wandering. Solid black circles represent the mean value of spectral wandering of several QEs for a given nanopillar height, while the error bars represent the calculated variance of each distribution, both extracted from time-resolved high-resolution spectral measurements (see Supplementary Section 4).

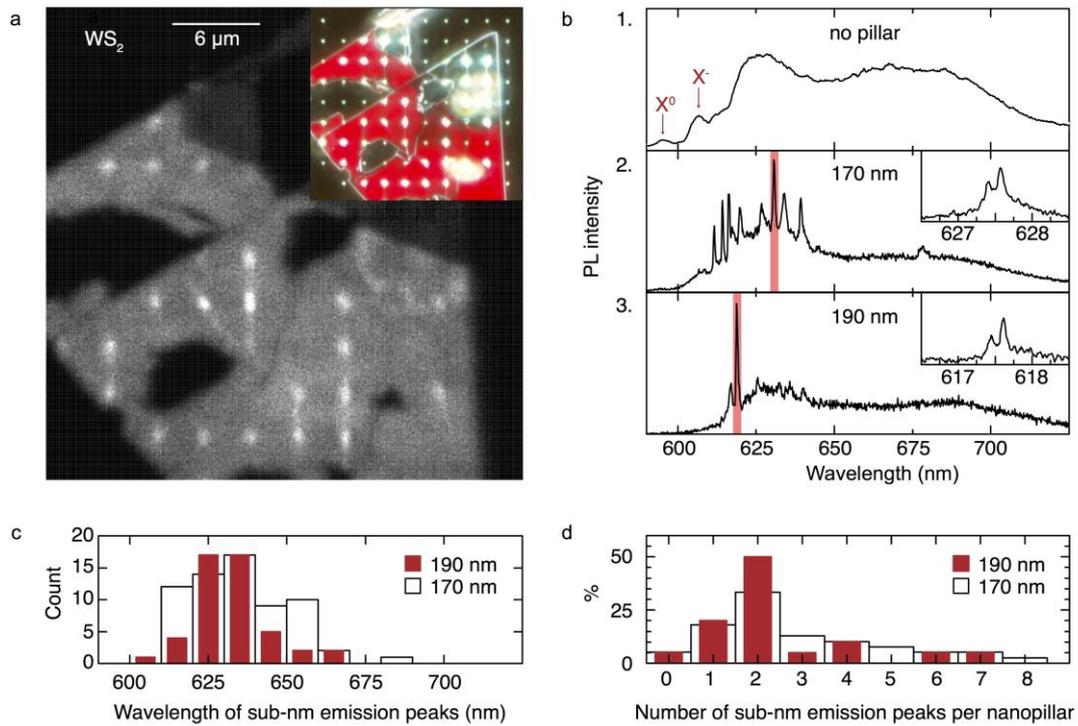

**Figure 3 | Creation of quantum-emitter arrays in 1L-WS$_2$. a,** Integrated PL intensity raster scan of a 1L-WS$_2$ flake deposited on top of a 3-µm spaced, 170-nm high nanopillar array, taken at 300-nW/µm$^2$, 532-nm laser excitation at 10 K. Colour-scale bar maximum is 18 kcounts/s. Inset: true-colour DFM image of the same area. The red region corresponds to the WS$_2$ monolayer. **b,** PL spectra of 1L-WS$_2$ at 10 K. Top panel 1 shows a spectrum taken from a flat region away from nanopillars. Red arrows indicate unbound monolayer neutral (X$^0$) and charged (X$^-$) excitons. Panels 2 and 3 show representative spectra of WS$_2$ on 170 and 190 nm nanopillars, respectively. Insets are high-resolution PL spectra of the red-highlighted spectral regions, showing the fine-structure splitting of the peaks. **c,** Distribution of the emission wavelengths measured for 1L-WS$_2$ QEs on 170 (black and white) and 190 nm (red) nanopillars. **d,** Distribution of the number of narrow emission lines observed per nanopillar for 1L-WS$_2$ QEs on 170 (black and white) and 190 nm (red) nanopillars.

# Supplementary Information

This supplementary information presents additional data regarding comments made in the main text and experimental observations.

**Table of Contents:**

## S1. Raman and PL material characterisation

Room temperature Raman and PL measurements are performed as discussed in the main text.

Figures S1a and b plot, respectively, the Raman and PL spectra of 1L-WS$_2$, as preliminarily identified by optical contrast, after transfer on the nanopillars. The Raman peaks at ~358 and ~419 cm$^{-1}$ correspond to the E' and A'$_1$ modes, respectively[1]. The separation between the two peaks is thickness-dependent[2], and increases with increasing number of layers[2]. Our value of ~61 cm$^{-1}$ indicates one-layer[2]. To further confirm this, we analyse its PL spectrum (Fig. S1b). A single peak at ~615 nm, corresponds to the neutral unbound exciton at the direct optical transition, a signature of 1L-WS$_2$[3]. We label this exciton X$^0$, following the notation used for TMDs by Ref. S4; elsewhere (e.g. in Ref. S3), the letter A is used, to distinguish it from a higher energy direct optical transition at ~520 nm (called "B") due to the spin-split valence band top.

Figures S1c,d (red lines) plot the Raman spectrum of 1L-WSe$_2$, as initially identified by optical contrast, after transfer on the nanopillars. For comparison we also measure in Figs S1c,d (blue lines) the spectrum of a 2L-WSe$_2$ flake on Si+285nm SiO$_2$, as identified by optical contrast. Figure S1c indicates that in the low frequency Raman region two additional peaks appear at ~17 cm$^{-1}$ and ~26 cm$^{-1}$ in 2L-WSe$_2$. The first peak, called C, is a shear mode caused by the relative motion of the layers, while the second peak is due to layer breathing modes[5,6] and can only appear in multi-layers. In Fig. S1d, red line, the peak at ~251 cm$^{-1}$, with full-width at half maximum (FWHM) ~2 cm$^{-1}$, is assigned to the convoluted A'$_1$+E' modes[1,2], degenerate in 1L-WSe$_2$[1,2], while the peak at ~262 cm$^{-1}$ belongs to the 2LA(M) mode. Due to the A'$_1$ and E' degeneracy, we do not use the separation between peak positions as fingerprint of the number of layers. In 2L-WSe$_2$ (Fig. S1d, blue line), the A$_{1g}$ and E$_g^1$ modes are degenerate at the same position of the peak in the 1L, ~251 cm$^{-1}$, and the peak has the same FWHM, ~2 cm$^{-1}$. The position of the 2LA(M) mode instead blue shifts to ~ 259 cm$^{-1}$, consistent with an increasing number of layers[2]. We also note the appearance of a peak at ~309 cm$^{-1}$, corresponding to the A$_{1g}^2$ mode that emerges only in multilayer WSe$_2$[1,2]. In order to further confirm the number of layers, we analyse the PL spectrum of 1L-WSe$_2$ (Fig. S1e, red line). We identify two features, one at ~750 nm (Fig. S1e, green line), corresponding to the neutral unbound exciton X$^0$ of 1L-WSe$_2$[3,4] and a second at ~770 nm (Fig. S1e, pink line), corresponding to the negatively charged unbound exciton X$^-$ of 1L-WSe$_2$[4], as validated by its redshift of ~20 nm (the positively charged exciton would redshift ~10 nm from X$^0$ due

to a smaller binding energy[4]). For reference, we compare the PL spectrum of 1L-WSe$_2$ to that of 2L-WSe$_2$ (Fig. S1e, blue line). The latter shows two components, at ~760 nm (orange line) and ~800 nm (purple line). The first corresponds to the direct optical transition, A[3], of 2L-WSe$_2$, while the second is due to its indirect optical transition, I[3].

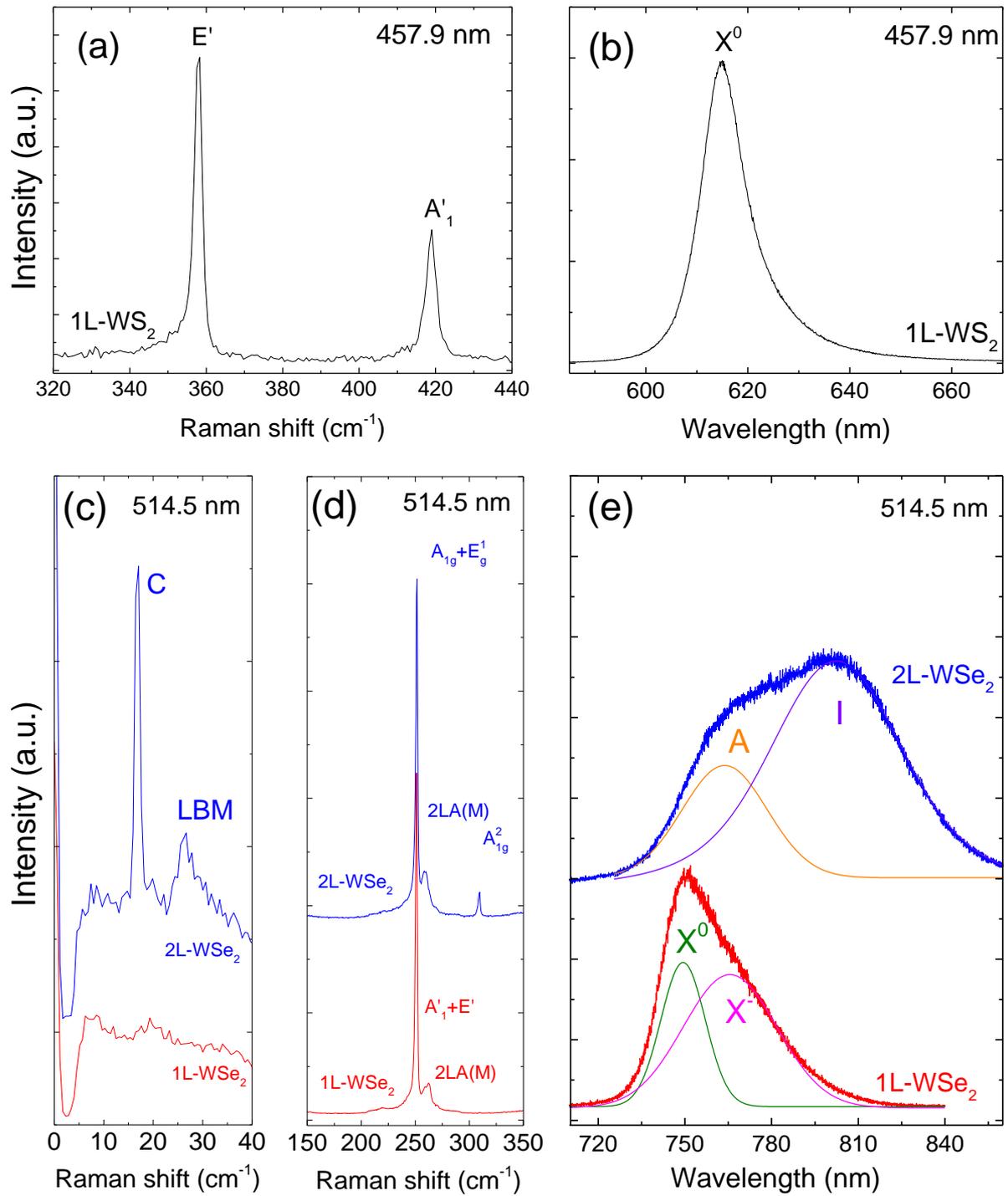

**Figure S1. Room temperature optical characterization of 1L-WS$_2$, 1L-WSe$_2$ and 2L-WSe$_2$.** (a) Raman and (b) PL spectra of 1L-WS$_2$ (black lines) after transfer on the patterned substrates; (c), (d) Raman and (e) PL spectra of 1L-WSe$_2$ (red lines) after transfer on the patterned substrates and of a reference sample of 2L-WSe$_2$ on Si/SiO$_2$ (blue line). The excitation wavelength is 514.5 nm for WSe$_2$ and 457 nm for WS$_2$.

## S2. Piercing of flakes

We use atomic force microscopy (AFM) scans to identify those flake sites that are either pierced or not pierced by the nanopillars. Figure S2a shows one such scan, where nanopillar sites are labelled 1-3 corresponding to: 1) a bare nanopillar outside the flake area; 2) a nanopillar that has pierced the flake; and 3) a nanopillar that has not pierced the flake. Fig. S2b indicates that the naked and pierced nanopillars (1 and 2) have a very similar profile, whereas nanopillar site 3 has approximately twice the width, which we assign to the flake draping over it. We correlate the dark field microscopy (DFM) images with AFM scans. As mentioned in the main text, non-pierced pillars appear as brighter spots in DFM due to a larger scattering area, compared to the dimmer pierced sites.

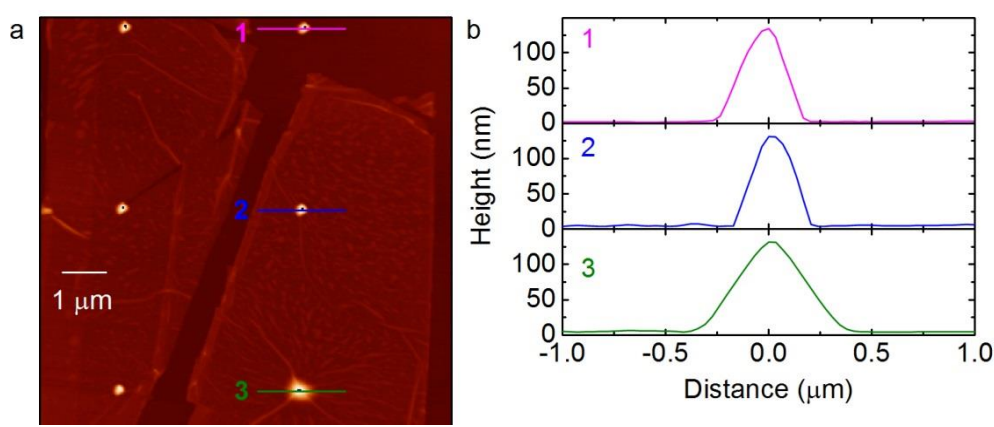

**Figure S2 | AFM characterisation of nanopillar sites. a,** AFM scan of a 1L-WSe$_2$ on 130-nm nanopillars. Colour scale maximum is 135 nm. Nanopillars are labelled: 1 (outside the flake, bare), 2 (under the flake, pierced) and 3 (under the flake, non-pierced). **b,** Height profiles across the lines over nanopillars 1 (pink), 2 (blue) and 3 (green). The full-width half-maximum measured for the nanopillars with no flake on top (1 and 2) are ~250 nm, while that of site 3 is ~500 nm, larger by as much as a factor ×2 due to the tenting of the flake over the nanopillar.

**S3. 1L-WSe$_2$ quantum emitter statistics**

We assess the effect of nanopillar height on the deterministic 1L-WSe$_2$ QEs by carrying out PL spectral measurements on 60, 130 and 190 nm nanopillars. The number of sub-nm peaks appearing per nanopillar and the measured spectral wandering show dependence on the nanopillar height, as displayed in Figs. 2e and 2f of the main text. As mentioned in the main text, the number of sub-nm peaks decreases for increasing height. Figure S3a shows representative examples of PL spectra measured at 10 K for each nanopillar height. Figure S3b plots the emission wavelength statistics collected for each nanopillar height, which are within the range of 730-820 nm. We carry out measurements on a total of over 80 nanopillar sites for the different nanopillar heights, observing no clear dependence of emission wavelength on nanopillar height, except a trend towards a narrower distribution of emission wavelength with increasing height. Figure S3c shows fine structure splitting values measured for each nanopillar height. These lay within the range 200-700 μeV, as discussed in the main text.

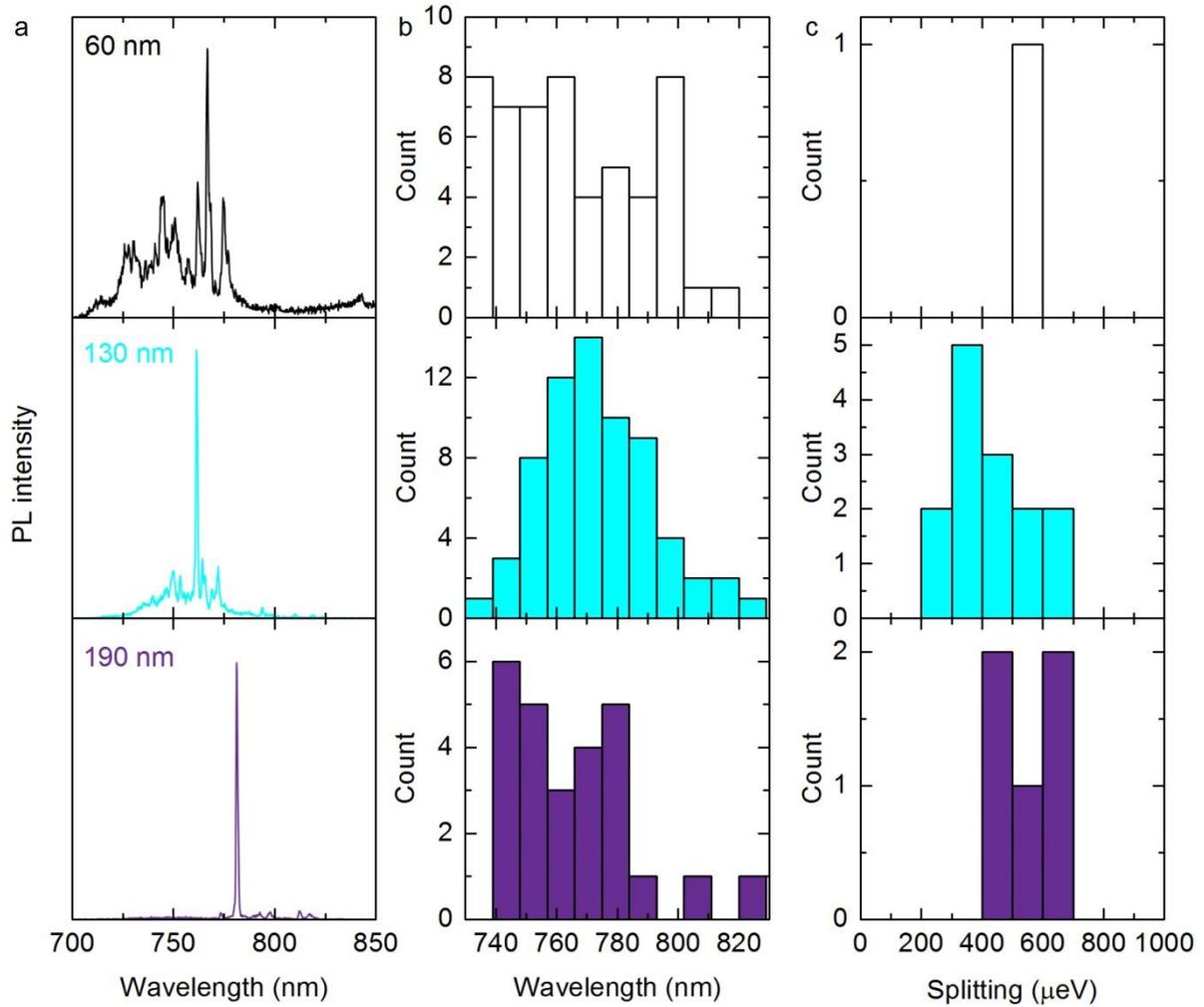

**Figure S3 | Effect of nanopillar height on 1L-WSe$_2$ QEs.** Statistics for QEs measured on nanopillars of heights 60 nm (black), 130 nm (cyan) and 190 nm (purple). **a,** Example PL spectra for the different nanopillar heights taken at 10 K. **b,** Histograms showing emission wavelength of all sub-nm lines studied. **c,** Fine structure splitting values of the 1L-WSe$_2$ QEs. We measure only one data point for the 60 nm nanopillars due to the QEs having linewidths in the range of 1 meV, generally greater than the fine structure splitting.

**S4. 1L-WSe$_2$ spectral wandering measurements**

Spectra of the sub-nm emission lines taken as a function of time show that QE spectral wandering decreases as the nanopillar height is increased. We show this in Fig. 2e of the main text. Figure S4 plots examples of time resolved spectra for each nanopillar height from which we extract the spectral wandering values, measured using a high-resolution spectrometer grating (1800 gr/mm) over 20 s.

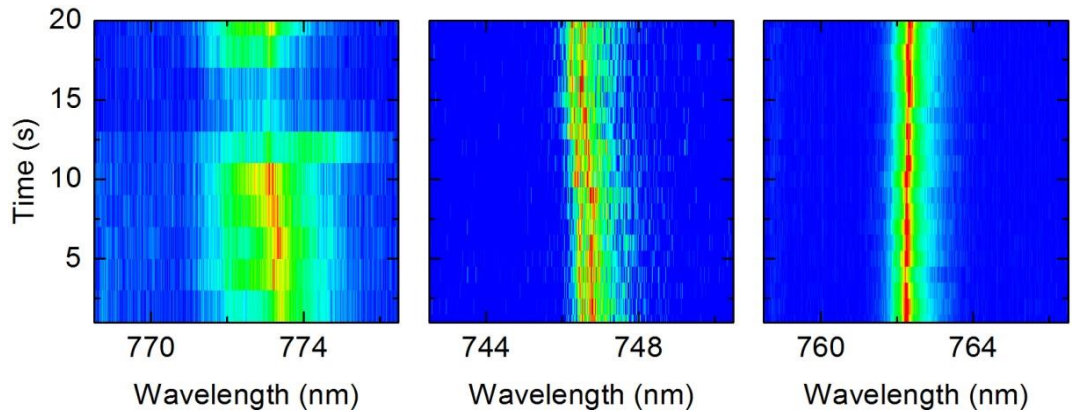

**Figure S4 | Spectral wandering of 1L-WSe$_2$ QEs as a function of nanopillar height.** High-resolution spectra at 10 K of QEs on different nanopillar heights taken over 20 s and showing a spectral window of 8 nm in each panel. The time resolution for each panel is 2, 1 and 1 s, respectively. Colour scale maxima are 45, 10 and 70 cts/s, respectively. Spectral wandering is reduced from ~1 to ~0.1 meV when going from 60 to 190-nm nanopillars.

## S5. Room temperature 1L-WSe$_2$ nanopillar PL measurements

We observe a redshift of the X$^0$ emission peak at room temperature (RT) at the nanopillar sites, as well as an increase in APD counts over the nanopillar region. Figure S5a shows a raster scan of integrated APD counts over one nanopillar site where several spectra have been taken across the dashed line. These spectra are shown in Fig. S5b, where a ~8 nm (15 meV) redshift can be seen. We measure redshifts from the X$^0$ peak ranging from 3 to 15 nm, with no clear nanopillar height dependence.

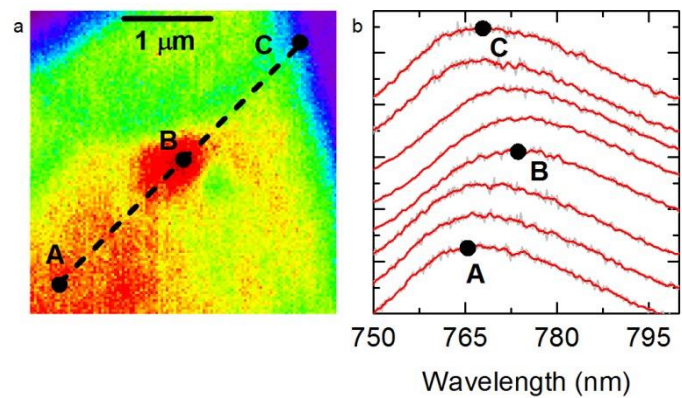

**Figure S5 | RT PL map of 1L-WSe$_2$ on 130 nm nanopillar substrate. a,** Integrated PL raster scan of one nanopillar site at RT, showing an increase in PL intensity at the nanopillar site (marked as B). Colour scale maximum is 590 kcounts/sec. The dashed line indicates where the spectra shown in the right panel are taken, with the first (A), central (B) and final (C) positions marked. **b,** RT spectra taken along the dashed line in a, showing an X$^0$ redshift of ~8 nm (15 meV).

## S6. 1L-WS$_2$ QE statistics

We carry out PL experiments on WS$_2$ using 60, 130, 170 and 190 nm nanopillars. We find QEs for nanopillars of heights 170 nm and 190 nm. Figures 3c and 3d of the main text show the effect of increased nanopillar height on the QE characteristics: a reduction in the spread of emission wavelengths and a reduced distribution of number of sub-nm lines appearing per nanopillar. Representative spectra of each nanopillar height are shown in Fig. S6a, exemplifying the reduction in peak number. Figure S6b shows spectral wandering for several QEs of both nanopillar heights, showing typically low values below 0.5 meV for 170 nm nanopillars. However, there are not enough statistics to distinguish a clear trend with nanopillar height. Figure S6c shows statistics collected for the fine structure splitting values measured. The values measured lie within the range 200 – 900 μeV, overlapping those observed in WSe$_2$ QEs.

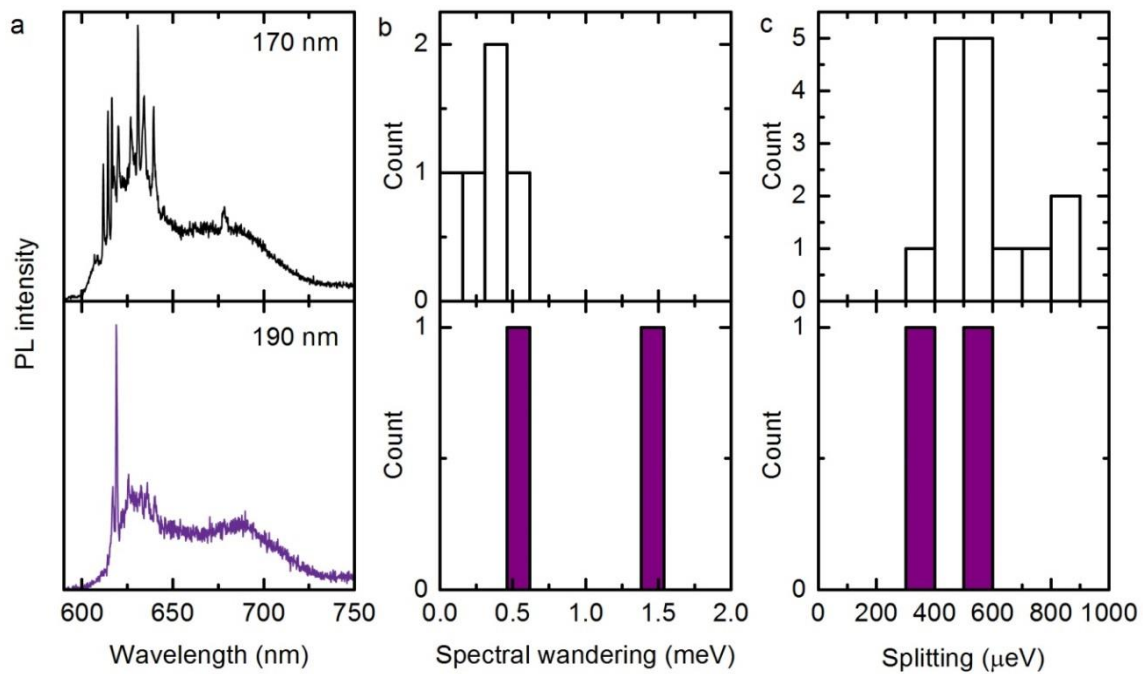

**Figure S6 | Effect of nanopillar height on 1L-WS$_2$ QEs.** PL Measurements taken at 10K. Data is shown in black for 170-nm nanopillars, and in purple for 190-nm nanopillars. **a,** Representative spectra taken at each nanopillar height, showing a reduction in the number of sub-nm lines with nanopillar height. **b,** Spectral wandering measurements taken as a function of nanopillar height, showing low spectral wandering but no clear trend as a function of height. **c,** Fine structure splitting values, previously unreported for WS$_2$ QEs, and lying in the same range as those values observed for WSe$_2$ QEs.

## S7. 1L-WS$_2$ on 60 and 130-nm nanopillars

We carry out PL measurements for 1L-WS$_2$ on 60 and 130 nm nanopillars at 10 K. As discussed in the main text, we detect QEs in 1L-WSe$_2$ on these substrates. However, we find no sub-nm lines for 1L-WS$_2$ for these pillar heights. Figure S7 shows example spectra for each height, taken at 10 K using the same (532 nm) laser excitation. The X$^0$ and X$^-$ bands are marked in grey for comparison. The broad red-shifted background seen in these scans is a usual feature that appears in 1L-WS$_2$ at low temperature, due to weakly localised emission bands[7].

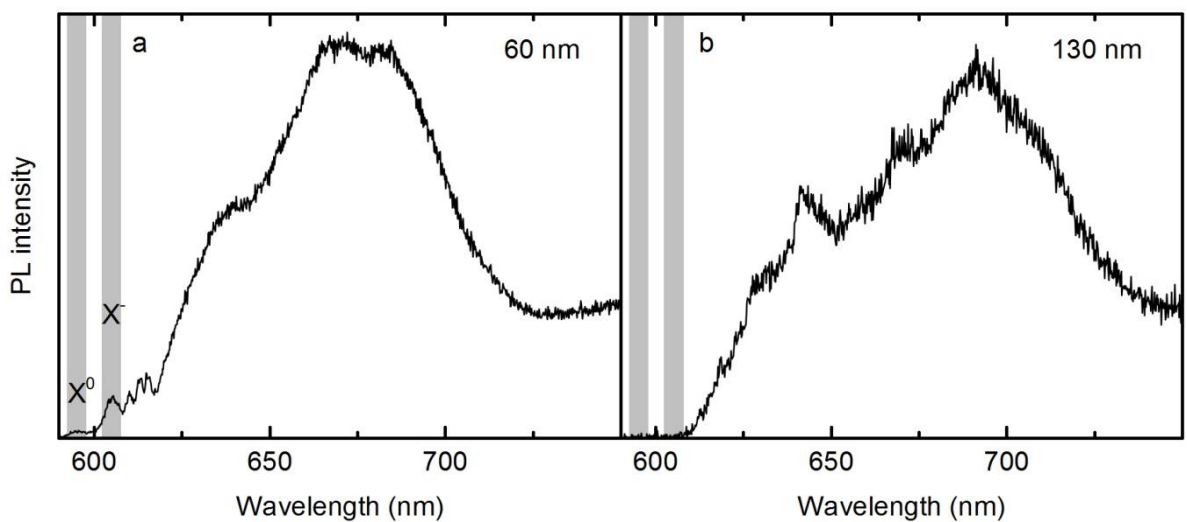

**Figure S7 | 1L-WS$_2$ on short nanopillars creating no quantum confinement.** Spectra taken at nanopillar sites on 1L-WS$_2$, at 10K on **a,** 60-nm and **b,** 130-nm nanopillars. No sub-nm peaks are observed. The broad shoulders correspond to the weakly localised emission observed in WS$_2$ at low temperatures, in regions both over nanopillars and in flats areas.

## S8. QE creation in 1L-WSe$_2$ using nanodiamonds

We deposit nanodiamonds, milled from bulk HPHT diamond (NaBond), of average diameter 100 nm onto SiO$_2$/Si substrates. We do this via a standard drop-casting technique, whereby we suspend the nanodiamonds in ethanol and deposit a drop onto the substrate using a pipette. The drop is left on the substrate for 1 minute and then washed with de-ionised water, leaving behind only those nanodiamonds stuck to the surface of the substrate. We then place 1L-WSe$_2$ flakes on them using the same viscoelastic technique as reported in the main text. These create similar protrusions or deformations in the flake as the nanopillars, but of varying sizes owing to the size and shape dispersion of the nanodiamonds. Figure S8a is an AFM scan of a 1L-WSe$_2$ flake on nanodiamonds. We take a height profile (shown in Fig. S8b) across the dashed line, where a nanodiamond is present under the flake. Figure S8c shows an integrated PL raster scan of the same sample taken at 10 K. The flake is highlighted by the white lines. There is an increase in PL intensity at the nanodiamonds site, similar to the effect seen with the nanopillars. Figure S8d shows a spectrum taken at this location, at 10 K and under 532 nm laser excitation, showing a sub-nm peak. About 20 such nanodiamonds-induced QEs were measured.

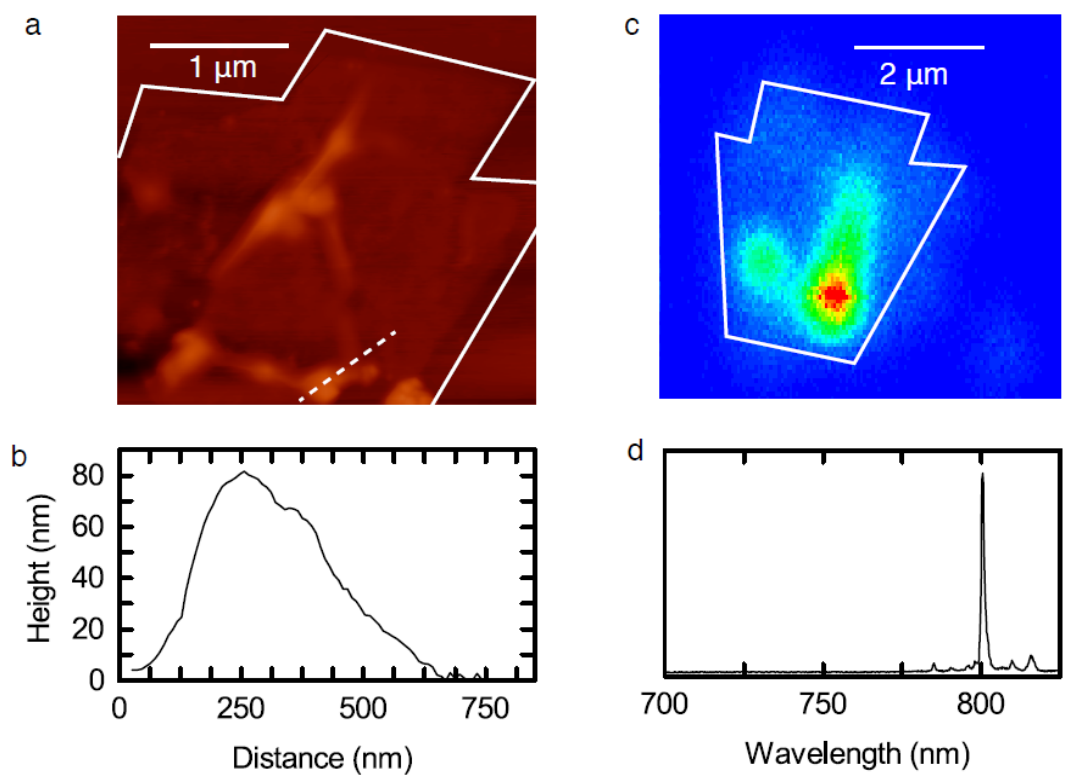

**Figure S8 | Nanodiamond substrates for QE creation. a,** AFM scan showing 1L-WSe$_2$ on a silica substrate with drop-casted nanodiamonds. The elevated regions of the flake are caused by nanodiamonds below. **b,** Height profile at the nanodiamond site taken along the white dashed line in a. **c,** Raster scan of integrated PL at 10 K of the same 1L-WSe$_2$ as shown in a. Colour scale maximum is 43 kcounts/s, and corresponds to the nanodiamond site in b. **d,** PL emission at 10K from the same nanodiamond region as shown in a and c.

**Supplementary References**

1. Terrones, H. *et al.* New first order Raman-active modes in few layered transition metal dichalcogenides. *Sci. Rep.* **4,** 4215 (2014).
2. Zhao, W. *et al.* Lattice dynamics in mono- and few-layer sheets of WS2 and WSe2. *Nanoscale* **5,** 9677–83 (2013).
3. Zhou, B. *et al.* Evolution of electronic structure in Atomically Thin Sheets of WS2 and WSe2. *ACS Nano* **7,** 791–797 (2013).
4. Jones, A. M. *et al.* Optical generation of excitonic valley coherence in monolayer WSe2. *Nat. Nanotechnol.* **8,** 634–638 (2013).
5. Zhang, X. *et al.* Raman spectroscopy of shear and layer breathing modes in multilayer MoS 2. *Phys. Rev. B* **87,** 115413 (2013).
6. Wu, J.-B. *et al.* Interface Coupling in Twisted Multilayer Graphene by Resonant Raman Spectroscopy of Layer Breathing Modes. *ACS Nano* **9,** 7440–7449 (2015).
7. Palacios-Berraquero, C. et al. Atomically thin quantum light-emitting diodes. *Nat. Commun.* 7, 12978 doi: 10.1038/ncomms12978 (2016).